\documentclass[12pt,preprint]{aastex}

\newfont{\boldit}{cmbxti7}

\newcommand{\PSbox}[3]{\mbox{\rule{0in}{#3}\includegraphics{#1}\hspace{#2}}}
\newcommand{\FigNum}[1]{\unitlength 1pt \begin{picture}(55,10)(-400,35)
                      \put(0,0){Figure #1}
                      \end{picture}}

\newcommand\approxgt{\mbox{$^{>}\hspace{-0.24cm}_{\sim}$}}

\def\x1608{{4U~1608$-$522}}

\def\cenx4{{Cen~X$-$4}}

\def\saxj1808{{SAX J1808.4$-$3658}}

\def\object3{CXOU~132619.7-472910.8}

\def\pid{\mbox{$P_{\rm id}$}}

\begin{document}

\title{Spectral Classification of Optical Counterparts to ROSAT All-Sky Survey X-ray Sources}

\author{Diana Dragomir,
Philippe Roy, Robert E. Rutledge\altaffilmark{1} }
\altaffiltext{1}{
Department of Physics, McGill University, 3600 rue University,
Montreal, QC,  H3A 2T8, Canada; dragomir, royp,
rutledge@physics.mcgill.ca
}

\begin{abstract}
Previous work statistically identified 5492 optical counterparts, with
\approxgt 90\% confidence, from among the $\approx$18,000 X-ray
sources appearing in the ROSAT All-Sky Survey Bright Source Catalog
(RASS/BSC).  Using low resolution spectra in the wavelength range 3700-7900\AA, we
present spectroscopic classifications for 195 of these counterparts which have
not previously been classified. Of these 195, we find 168 individual
stars of F, G, K or M type, 6 individual stars of unknown type, 6
double stars, 6 AGN or galaxies and 7 unclassifiable objects; the spectra of the 2
remaining objects were saturated.
\end{abstract}


\section{Introduction}

The X-ray sources in the {\em ROSAT} All-Sky Survey/Bright Source
Catalog \citep{bsc}, are the brightest $\approx$18,000 soft sources (0.1-2.4 keV)
in the sky.  There exists no comprehensive catalog which
identifies counterparts in off-bands (optical, radio or infra-red) for
all of them in a statistically rigorous fashion: that is, in which the
probability of association between each X-ray source and off-band
counterpart is given.

In previous work \citep{Rutledge00}, 5492 of the $\approx$18,000 RASS/BSC
sources were found to have optical counterparts -- sources which were
uniquely associated with the RASS/BSC X-ray sources, with a
probability of unique association $\pid>$90\%. That is, each optical
counterpart has $<$10\% probability of not being the sole optical
counterpart of the specified RASS/BSC source, due to their
statistically improbable close proximity to the X-ray source, and high
USNO-A2 \citep{USNO} $B$-band optical brightness, in comparison with background
fields. Most of these 5492 optical counterparts had been previously
classified, but many have not.  We have therefore undertaken a program
of optical spectral classification of these identified counterparts of
RASS/BSC sources.

In this paper, we present optical spectral classification of 195
optical counterparts to RASS/BSC sources, selected from among the 5492
RASS/BSC X-ray sources with optical counterparts with \pid$\geq0.9$,
but which did not have source classifications at the time of the
observations.  This group of sources has an average value of
\pid=0.957; we expect, on average 187 of the 195 RASS/BSC sources
herein classified to be correctly identified. All 195 sources and their
corresponding probabilities of unique association, RASS countrates, USNO-A2
brightness and classifications are listed in Table \ref{tab:obs}. We have
found that 18 of those sources had been previously classified. These sources
and their corresponding previous classifications are listed in Table \ref{tab:prev}.
Most (117) sources necessitate comments regarding their classification. These comments 
are enumerated in Table \ref{tab:comments}. The first column of each of these tables
contains a running number in order to facilitate the assignation of a previous classification
or comment to the respective ID information.

In \S~\ref{sec:obs}, we describe the observations and analyses,
reducing the observational spectra to flattened spectra.  In
\S~\ref{sec:classification}, we briefly describe the method of
classification for each source class considered (that is, AGN/Galaxy,
F star, K star, etc.). In \S~\ref{sec:con} we discuss these
classifications and conclude.

\section{Observations and Analysis}
\label{sec:obs}

The optical spectra were obtained using the Palomar 60-inch echelle
spectrograph \citep{mccarthy88}, in combination with a TI 800$\times$800
pixel CCD. A 1.4 arcsec slit projecting to 2 pixels was used, with spectra 
covering the range 3700-7900 ${\AA}$ at a resolution $R=\lambda/\Delta \lambda=40,000$ 
per pixel, for an effective resolution of $R=20,000$. Calibration spectra were taken
from a Th-Ar hollow cathode tube. The spectra were taken during seven nights,
from April 28 to May 4, 2003.

\subsection{Data Reduction}

The reduction of the echelle spectra was performed using the NOAO (National
Optical Astronomy Observatories)
package in IRAF (Image Reduction and Analysis Facility). The steps
taken are as follows:

\begin{itemize}
\item Trimming of edge defects from all exposures;
\item Combination and averaging of the bias and flat-field exposures;
\item Bias-correction of the flat-field and object exposures;
\item Division of the object exposures by the averaged flat-field exposure;
\item Extraction of the Th-Ar calibration exposure and identification of the Th-Ar emission lines;
\item Dispersion calibration of the object exposures and extraction of the spectra;
\item Normalization to the continuum.
\end{itemize}

The reduction procedure mainly followed the steps outlined in 
A User's Guide to Reducing Echelle Spectra with IRAF \citep{Iraf}, with a few changes.

The overscan correction was omitted, since we did not perform a flux
calibration.   A bias correction was nonetheless performed on the
combined and averaged flat-field exposure as well as on the object
exposures.

To preserve the profile shape of the flat-field exposure, the latter
was not normalized. The object frames were divided by the flat-field
with the {\tt ccdproc} task. Then the Th-Ar calibration frames were
extracted and a dispersion solution was found with the {\tt
ecidentify} task.

Using the {\tt doecslit} task in the echelle subpackage, the apertures
of the flat-fielded object frames were centered and resized, and the
object spectra were dispersion calibrated and extracted. Finally,
normalization to the continuum was performed using the {\tt continuum}
task in the {\tt noao.onedspec} package.

\section{Classification}
\label{sec:classification}

The orders showing the H${\alpha}$ Balmer line, the Li I line (6103
{\AA}), the Ca I line (6122 {\AA}) and Fe I line (6136 {\AA}), the Na
I doublet lines, the Mg I triplet lines, the Ca I (4227{\AA}) line,
and the Ca II H \& K lines have been used for the classification of
most objects. For some of the objects classified as M stars, the order
showing the TiO lines (7088 {\AA} and 7125 {\AA}) has also been used
in the classification. For a number of objects, the orders showing the
Ca I (4227{\AA}) line and the Ca II H \& K lines have not been used in
the classification. 
The accuracy of the wavelength calibration
for those orders for those objects is too low and thus an analysis of
the spectral features observed in these orders would not be
sufficiently reliable. This effect occurs in the blue orders and it is
due to the fact that the edges of those orders have a lower signal
than their central portions.

We have included seven figures showing a sample representing the most
relevant (i.e. the most useful for the classification) of these 
spectroscopic features for selected objects.

\subsection{Classification criteria for individual objects}

In this section, we describe the basis for classification of each
source type used in Table \ref{tab:obs}.  Note that lines are absorption
lines unless specified otherwise.

{\bf AGN/Galaxy:} H${\alpha}$ Balmer line redshifted by 3 ${\AA}$ or
more (indicating high recessional velocity); redshifted metal lines
and Ca I lines; generally weak lines.  Cross-check with 2MASS image
database shows object is spatially resolved.

{\bf Star:} noisy spectrum; some lines difficult to identify and their
strength difficult to estimate due to low signal-to-noise ratio; lines
are in absorption and deep and narrow.

{\bf F star:} strong H Balmer lines; neutral metal and Na I D lines
(Na I D lines are shallower and narrower than in K stars); Ca I line
at 4227 ${\AA}$; Ca II H and K lines (strongest absorption features)
with negligible core emission.

{\bf F/G star:} strong H Balmer lines; neutral metal and Na I D lines
(Na I D lines are shallower and/or narrower than in K stars); Ca I
line at 4227 ${\AA}$; Ca II H and K lines (strongest absorption
features) with weak core emission. Difficult to conclude spectral type
from line strength; these objects are most likely to be late F stars
or early G stars.

{\bf G star:} strong H Balmer lines; neutral metal and Na I D lines
(Na I D lines are deep but narrower than in K stars); Ca I line at
4227 ${\AA}$; Ca II H and K lines (strongest absorption features),
usually with core emission.

{\bf G/K star:} H Balmer lines; neutral metal and Na I D lines; Ca I
line at 4227 ${\AA}$; Ca II H and K lines with emission (in some cases
the Ca II H and K lines are difficult to find due to low
signal-to-noise ratio).  Difficult to conclude spectral type from line
strength, especially if Ca II H and K lines are difficult to find.

{\bf F/G/K star:} noisy spectrum; H Balmer lines; neutral metal and Na
I D lines; Ca I line at 4227 ${\AA}$; Ca II H and K lines; difficult
to estimate strength of lines due to low signal-to-noise ratio; a more
specific classification would be unreliable.

{\bf K star:} H Balmer lines; neutral metal and strong Na I D lines;
Ca I line at 4227 ${\AA}$ (strongest absorption feature); Ca II H and
K lines with core emission.

{\bf K/M star:} H Balmer lines (weak if in absorption); neutral metal
and strong Na I D lines; Ca I line at 4227 ${\AA}$; Ca II H and K
lines with core emission; occasionally some weak TiO lines.  Difficult
to conclude spectral type from line strength; these objects are likely
to be late K stars or early M stars.

{\bf G/K/M star:} noisy spectrum; H Balmer lines (not strong enough
for a F star); neutral metal and rather strong Na I D lines; Ca I line
at 4227 ${\AA}$; Ca II H and K lines; difficult to deduce strength of
most lines because of noise; a more specific classification would be
unreliable.

{\bf M star:} weak H Balmer lines; neutral metal (Mg I triplet
especially strong) and very strong Na I D lines; Ca I line at 4227
${\AA}$; Ca II H and K lines with core emission; conspicuous TiO
lines.

{\bf Unclassifiable:} very noisy spectrum; featureless spectrum or few
weak lines (insufficient for reliable classification).

{\bf Double:} cross-check with 2MASS image database showed two objects
less than 15'' apart and of approximately same magnitude; either one of the two or both
could be the optical counterpart; both objects are
classified separately, using the criteria described above.

{\bf Saturated:} saturated spectrum.

For all objects classified as stars, a cross-check with the 2MASS
image database showed object has a narrow point spread function.

Sources with H Balmer lines in emission and/or strong Ca II H and/or K 
core emission and/or with lines that
exhibit P Cygni profile are classified as coronally
active.

\section{Discussion and Conclusions}
\label{sec:con}

We reduced the low resolution echelle optical spectra of 195
counterparts to X-Ray sources found in the ROSAT All-Sky Survey Bright
Source Catalog using the NOAO package in IRAF.  We classified these
counterparts and found 168 individual stars of F, G, K or M type, 6
individual stars of unknown type, 6 double stars and 6 AGN or
galaxies. Seven of the analysed spectra were found to be unclassifiable
and two were saturated.

We suspect the unclassifiable objects are objects such as AGN,
galaxies or white dwarves, since there were no strong lines typical of
main type stars in their optical spectra.  Furthermore, the previous
classifications of one of those objects (121506.7+331129 (\# 33)) found it to
be a AGN, and those of two other of those objects (182030.0+580437 (\# 176) and
184739.3+015732 (\# 192)) found them to be white dwarves, thus upholding our
hypothesis.

The spectra of some 12 objects exhibit some or most of the spectroscopic
features used for the classification as double lines. This indicates that
at least some of these objects may be binary systems.  
The spectra of the K/M star of 125534.6+270355 (\# 44) shows that 
most of the lines used for the spectral classification are double. Furthermore,
\# 44 was classified as a Double (with the second object being an unclassified 
star), thus increasing the probability of it being a binary system.
In general, the spectral resolution used for the observations was insufficient 
for confident detection of such systems.

We also found a possible discrepancy between our classification of
optical counterpart 171632.8+430221 (\# 128) and its previous classification
\citep{Hamburg03}. We classified this counterpart as a star of type
K or M, whereas it has previously been classified as a galaxy.  We
cannot rule out that the object analysed by us is different from that
analysed by Zickgraf et al. (2003).

The optical counterparts analysed deserve further observation to
refine classification.  However, the aim of our analysis is to provide
a general idea of the nature of the optical counterparts we studied.
We were successful for the majority of the counterparts and in the
case of most of the objects classified as stars, we were able to find
their approximate type. In some cases, indicators of coronal activity
in stars is also observed.  These results can be used for selective
further investigation of these counterparts.

\acknowledgments

RER acknowledges support from NSERC through the Discovery program.

\clearpage

\begin{deluxetable}{cccccc}
\tabletypesize{\scriptsize}
\tablecaption{\label {tab:obs} Table of Sources}
\tablewidth{14cm}
\tablehead{
\colhead{\#} &
\colhead{Source Name} &
\colhead{$P_{\rm id}$} &
\colhead{I (RASS countrate)} &
\colhead{$B_{\rm USNO-A2}$} &
\colhead{Source Class}  \\
\colhead{} &
\colhead{(1 RXS J)} &
\colhead{} &
\colhead{(c/sec) ({$1\sigma$})}  &
\colhead{(mag)} &
\colhead{} 
}

\startdata
1 & 103759.7+615310 & 0.967 & 0.0934 (0.0157) & 12.6 & AGN/Galaxy \\ 
2 & 103858.4+711254 & 0.976 & 0.1341 (0.0225) & 11.9 & G/K * \\
3 & 103945.8+654529 & 0.917 & 0.1574 (0.0171) & 13.0 & M *+ \\
4 & 104038.7+373233 & 0.924 & 0.1710 (0.0221) & 13.5 & Unclassifiable  \\
5 & 104809.7+354946 & 0.953 & 0.0888 (0.0183) & 12.1 & G/K *  \\
6 & 105026.4+503744 & 0.905 & 0.0998 (0.0157) & 12.7 & G/K *  \\
7 & 105159.4-090433 & 0.968 & 0.0948 (0.0197) & 11.9 & G/K *  \\
8 & 105541.8+424603 & 0.976 & 0.1767 (0.0218) & 11.0 & G/K *   \\
9 & 105634.4+583535 & 0.905 & 0.0973 (0.0174) & 12.7 & G/K *  \\
10 & 105706.3-010109 & 0.948 & 0.1210 (0.0189) & 10.9 & ** (G * and G/K *)  \\
11 & 105711.2+285626 & 0.987 & 0.0539 (0.0148) & 10.4 & K *  \\
12 & 110320.5+355900 & 0.999 & 0.1670 (0.0241) & 0.0 & M *  \\
13 & 110830.8+011742 & 0.995 & 0.0726 (0.0148) & 11.0 & ** (both G*)  \\
14 & 111232.8+354845 & 0.959 & 0.3203 (0.0446) & 11.9 & G * \\ 
15 & 111255.8+073325 & 0.955 & 0.0532 (0.0129) & 12.1 & G/K *  \\
16 & 111612.6+494208 & 0.931 & 0.1287 (0.0209) & 13.1 & K/M *+  \\
17 & 111650.2+445424 & 0.976 & 0.0935 (0.0177) & 11.6 & G/K *  \\
18 & 111712.4+293408 & 0.925 & 0.0836 (0.0217) & 9.9 & F/G *  \\
19 & 111900.0+643842 & 0.981 & 0.0724 (0.0126) & 11.7 & F/G *  \\
20 & 112152.3+042021 & 0.963 & 0.0510 (0.0132) & 10.8 & F/G *  \\
21 & 113300.5+031156 & 0.972 & 0.0673 (0.0151) & 12.1 & F/G/K *  \\
22 & 113316.5+134427 & 0.953 & 0.0788 (0.0158) & 12.9 & K/M *+  \\
23 & 113336.8+075131 & 0.946 & 0.1108 (0.0186) & 12.5 & F/G *+   \\
24 & 113352.8+361331 & 0.941 & 0.1096 (0.0219) & 12.9 & K/M *+   \\
25 & 113533.7+825920 & 0.977 & 0.2383 (0.0209) & 12.0 & K *+  \\
26 & 113858.9+421957 & 0.954 & 0.1070 (0.0243) & 9.8 & K *  \\
27 & 114839.6+231136 & 0.982 & 0.1017 (0.0163) & 11.8 & F/G *  \\
28 & 115107.6+351617 & 0.948 & 0.0687 (0.0155) & 12.0 & M *  \\ 
29 & 115651.8+082726 & 0.969 & 0.1935 (0.0246) & 12.3 & K/M *+   \\
30 & 115938.3+560301 & 0.965 & 0.0885 (0.0166) & 12.8 & G/K *+   \\
31 & 120732.6+471447 & 0.959 & 0.1234 (0.0235) & 12.4 & G/K *  \\
32 & 121215.7+685306 & 0.987 & 0.1514 (0.0164) & 11.4 & G/K *+   \\
33 & 121506.7+331129 & 0.985 & 0.5210 (0.0740) & 7.8 & Unclassifiable \\ 
34 & 121754.3+105647 & 0.933 & 0.0535 (0.0151) & 12.8 & G/K *  \\
35 & 121824.3+701748 & 0.980 & 0.0871 (0.0126) & 11.7 & G/K *  \\
36 & 121906.1+182119 & 0.972 & 0.0954 (0.0177) & 12.1 & G/K *   \\
37 & 122212.6+731453 & 0.952 & 0.2364 (0.0179) & 9.1 & G *+  \\ 
38 & 122308.2+452801 & 0.948 & 0.0970 (0.0180) & 12.5 & G/K *+   \\
39 & 122549.8+094550 & 0.981 & 0.0897 (0.0169) & 11.9 & G/K *  \\
40 & 123858.2+584417 & 0.972 & 0.0872 (0.0155) & 12.0 & F/G *  \\
41 & 123915.4+034216 & 0.937 & 0.0628 (0.0132) & 12.9 & G/K *  \\
42 & 124420.6+461916 & 0.984 & 0.0529 (0.0124) & 11.7 & G/K *  \\
43 & 124848.1+245031 & 0.983 & 0.1332 (0.0206) & 11.8 & G *  \\
44 & 125534.6+270355 & 0.983 & 0.0834 (0.0167) & 11.5 & ** (K/M * and *+)  \\ 
\hline
\tablebreak
45 & 130017.6-083929 & 0.987 & 0.1382 (0.0245) & 11.0 & F/G *  \\
46 & 130038.5+182220 & 0.967 & 0.0704 (0.0142) & 11.8 & F/G *  \\
47 & 130205.5+151114 & 0.948 & 0.0521 (0.0142) & 11.7 & G/K *  \\
48 & 130346.2+283729 & 0.970 & 0.0648 (0.0135) & 12.0 & G *+  \\ 
49 & 130349.5-050932 & 0.997 & 0.1699 (0.0277) & 8.7 & G/K *  \\
50 & 131152.9+275242 & 0.998 & 0.2550 (0.0370) & 0.0 & Saturated \\
51 & 132117.1+210116 & 0.925 & 0.0560 (0.0162) & 12.8 & *  \\
52 & 132641.4+022230 & 0.941 & 0.0558 (0.0163) & 12.8 & F/G *  \\
53 & 132713.0+455826 & 0.969 & 0.1569 (0.0185) & 12.4 & K/M *+   \\
54 & 132906.4+062743 & 0.985 & 0.0785 (0.0192) & 11.6 & G/K *  \\
55 & 132931.8+514214 & 0.908 & 0.0545 (0.0127) & 13.5 & F/G/K *+   \\
56 & 133443.7-082036 & 0.984 & 1.0610 (0.0600) & 10.9 & K *+   \\ 
57 & 133740.0+320930 & 0.970 & 0.0877 (0.0144) & 12.2 & K/M *+   \\
58 & 134426.3+441624 & 0.967 & 0.1119 (0.0157) & 12.7 & AGN/Galaxy  \\
59 & 134838.5+712552 & 0.962 & 0.1164 (0.0141) & 12.9 & K/M *+   \\
60 & 135441.1+155624 & 0.955 & 0.0560 (0.0151) & 12.2 & G/K *  \\
61 & 135510.3+160837 & 0.967 & 0.1458 (0.0222) & 12.7 & K/M *+   \\
62 & 135728.9+591107 & 0.970 & 0.0598 (0.0119) & 12.0 & G/K *  \\
63 & 135902.2+591754 & 0.950 & 0.0670 (0.0134) & 12.6 & K/M *+   \\
64 & 141357.5+301320 & 0.921 & 0.0522 (0.0131) & 9.5 & K *  \\
65 & 141508.0+124249 & 0.930 & 0.0761 (0.0165) & 12.6 & *  \\
66 & 141630.7+265523 & 0.975 & 0.1363 (0.0190) & 11.7 & G *+   \\
67 & 142019.9+275851 & 0.965 & 0.0578 (0.0122) & 12.4 & G/K *  \\
68 & 142057.5+513733 & 0.967 & 0.0679 (0.0116) & 12.9 & K/M *  \\
69 & 142408.3+210535 & 0.938 & 0.0939 (0.0162) & 11.6 & AGN/Galaxy  \\
70 & 143440.3+294441 & 0.998 & 0.1706 (0.0225) & 10.2 & Saturated  \\ 
71 & 143453.4+595146 & 0.965 & 0.1252 (0.0210) & 12.4 & Unclassifiable  \\
72 & 143554.1+173757 & 0.967 & 0.0535 (0.0129) & 12.3 & G/K *  \\
73 & 143600.5+094453 & 0.996 & 0.1489 (0.0239) & 8.7 & K *  \\
74 & 143638.2+584309 & 0.969 & 0.1034 (0.0167) & 12.5 & *  \\
75 & 143854.8+330022 & 0.970 & 0.0546 (0.0132) & 12.3 & G/K *  \\
76 & 144145.0+423124 & 0.965 & 0.0806 (0.0127) & 12.5 & *  \\
77 & 144308.7+172042 & 0.924 & 0.1054 (0.0184) & 12.6 & F/G *  \\
78 & 144952.3+420615 & 0.948 & 0.0631 (0.0109) & 12.2 & G/K *  \\
79 & 145502.5+144210 & 0.933 & 0.0909 (0.0193) & 12.7 & K *+   \\
80 & 145755.4+193623 & 0.958 & 0.0556 (0.0146) & 12.8 & G/K *  \\
81 & 150507.5+573036 & 0.942 & 0.0662 (0.0114) & 12.6 & K/M *  \\
82 & 151145.7+101421 & 0.945 & 0.0773 (0.0196) & 13.1 & K/M *+   \\
83 & 151750.8+050615 & 0.949 & 0.1071 (0.0194) & 11.7 & AGN/Galaxy  \\ 
84 & 151948.2+070754 & 0.920 & 0.1928 (0.0257) & 12.2 & K *   \\
85 & 152244.2+161550 & 0.993 & 0.4247 (0.0408) & 8.8 & G *  \\
86 & 152553.0+614216 & 0.927 & 0.0601 (0.0095) & 0.0 & *  \\
87 & 152757.4+651533 & 0.987 & 0.1420 (0.0133) & 11.4 & G/K *  \\
88 & 153726.3+362343 & 0.940 & 0.0658 (0.0159) & 12.4 & G/K *  \\
\hline
\tablebreak
89 & 154058.3+402657 & 0.974 & 0.0937 (0.0166) & 11.8 & G/K *  \\
90 & 154150.8+310031 & 0.945 & 0.1612 (0.0274) & 13.0 & AGN/Galaxy  \\
91 & 154355.8+132555 & 0.949 & 0.0539 (0.0119) & 12.4 & K *  \\
92 & 154457.5+793054 & 0.941 & 0.0596 (0.0097) & 12.3 & Unclassifiable  \\
93 & 154623.6+440318 & 0.924 & 0.0818 (0.0159) & 13.4 & G/K/M *+   \\
94 & 154718.8+631101 & 0.957 & 0.0863 (0.0119) & 13.0 & G/K/M *+   \\
95 & 154809.0+161207 & 0.946 & 0.0639 (0.0140) & 12.9 & M *+   \\
96 & 154925.0+460822 & 0.999 & 0.1368 (0.0184) & 0.0 & ** (G/K + * and K/M *)  \\
97 & 155752.5+774738 & 0.982 & 0.0650 (0.0097) & 11.6 & F/G *  \\
98 & 155842.6+323047 & 0.928 & 0.0548 (0.0138) & 12.9 & M *+    \\
99 & 155947.5+440358 & 0.929 & 0.1827 (0.0228) & 13.5 & M *+   \\
100 & 160157.0+182514 & 0.952 & 0.0942 (0.0142) & 12.9 & M *+   \\
101 & 160248.3+252031 & 0.988 & 0.1967 (0.0191) & 0.0 & K *+   \\
102 & 160438.3+702212 & 0.966 & 0.0819 (0.0095) & 12.5 & K/M *+   \\
103 & 160524.8-085549 & 0.918 & 0.0510 (0.0119) & 13.3 & G/K *+   \\
104 & 160704.4+023823 & 0.993 & 0.2030 (0.0229) & 11.0 & G/K *  \\
105 & 161643.2+671430 & 0.901 & 0.0599 (0.0075) & 10.3 & M *  \\
106 & 161712.7+773345 & 0.924 & 0.0884 (0.0104) & 13.7 & K/M *+   \\
107 & 162506.5+300218 & 0.982 & 0.2195 (0.0201) & 11.4 & K *+   \\
108 & 162526.8+572730 & 0.963 & 0.1860 (0.0139) & 13.0 & AGN/Galaxy  \\
109 & 162708.9+662620 & 0.960 & 0.0621 (0.0074) & 12.5 & G/K *+   \\
110 & 162946.1+281034 & 0.958 & 0.0923 (0.0143) & 12.0 & F/G *  \\
111 & 163259.1+504433 & 0.967 & 0.0915 (0.0153) & 12.5 & K *+   \\
112 & 163739.5+221104 & 0.970 & 0.1483 (0.0221) & 11.7 & K *  \\
113 & 163741.2+291946 & 0.965 & 0.0791 (0.0145) & 12.7 & K/M *+   \\
114 & 164212.9+655300 & 0.976 & 0.0891 (0.0082) & 12.3 & G/K *+   \\
115 & 165315.4+701554 & 0.964 & 0.0908 (0.0072) & 12.5 & F/G *  \\
116 & 165357.6+073450 & 0.969 & 0.0662 (0.0132) & 12.2 & G/K *  \\
117 & 165601.2+650807 & 0.990 & 0.4058 (0.0133) & 10.6 & F *  \\ 
118 & 165733.5+593156 & 0.967 & 0.0544 (0.0084) & 12.2 & F/G *  \\
119 & 165909.5+205807 & 0.927 & 0.3880 (0.0261) & 13.6 & M *+   \\
120 & 170157.5+073329 & 0.947 & 0.0581 (0.0126) & 13.1 & F/G *  \\
121 & 170248.8+471258 & 0.943 & 0.2374 (0.0203) & 12.1 & G/K *+   \\
122 & 170420.7+392909 & 0.932 & 0.0621 (0.0097) & 12.3 & G/K *+   \\
123 & 170757.4+291922 & 0.911 & 0.0891 (0.0127) & 12.9 & G/K *+   \\
124 & 171017.5+632135 & 0.974 & 0.0520 (0.0056) & 11.9 & G/K *  \\
125 & 171206.0+454025 & 0.966 & 0.0581 (0.0088) & 0.0 & M *  \\
126 & 171331.0+232021 & 0.987 & 0.1021 (0.0136) & 11.3 & G *  \\
127 & 171355.0+455955 & 0.977 & 0.0502 (0.0073) & 11.9 & G/K *  \\
128 & 171632.8+430221 & 0.984 & 0.1323 (0.0139) & 11.7 & K/M * \\ 
129 & 171800.1+212816 & 0.989 & 0.0534 (0.0111) & 10.8 & F/G *  \\
130 & 172103.4+114237 & 0.988 & 0.1549 (0.0180) & 0.0 & ** (G/K * and M *+ )  \\
131 & 172128.0+084446 & 0.967 & 0.0564 (0.0129) & 12.4 & G/K *+   \\
132 & 172157.8+574913 & 0.917 & 0.0676 (0.0073) & 12.5 & G/K *+   \\
\hline
\tablebreak
133 & 172325.9+193119 & 0.983 & 0.0630 (0.0121) & 11.9 & K *  \\
134 & 172339.8+352757 & 0.970 & 0.0654 (0.0107) & 12.5 & F/G *   \\
135 & 172413.5+402616 & 0.972 & 0.0552 (0.0094) & 12.3 & K/M *  \\
136 & 172534.0+160923 & 0.970 & 0.3358 (0.0377) & 12.8 & G/K *+   \\
137 & 172812.2+723923 & 0.962 & 0.0657 (0.0063) & 12.9 & K *+   \\
138 & 172927.2+352402 & 0.944 & 0.0842 (0.0112) & 13.2 & G/K *+   \\
139 & 173039.4+785554 & 0.973 & 0.1114 (0.0087) & 12.2 & K *+   \\
140 & 173103.4+281510 & 0.994 & 0.2869 (0.0215) & 0.0 & G/K *  \\
141 & 173306.7+091433 & 0.979 & 0.0597 (0.0132) & 10.5 & K/M *  \\
142 & 173614.7+650229 & 0.985 & 0.1022 (0.0056) & 12.0 & K *  \\
143 & 173734.1+414618 & 0.965 & 0.1123 (0.0112) & 12.7 & G/K *+   \\
144 & 173748.1+215925 & 0.939 & 0.0566 (0.0109) & 13.0 & F/G/K *+   \\
145 & 173855.7+263400 & 0.919 & 0.1780 (0.0175) & 13.6 & M *+   \\
146 & 174120.9+084312 & 0.953 & 0.0677 (0.0139) & 11.8 & G/K *+   \\
147 & 174311.5+334950 & 0.999 & 0.1235 (0.0140) & 0.0 & *+  \\
148 & 174411.0+720304 & 0.985 & 0.1424 (0.0069) & 11.9 & K *+   \\
149 & 174432.1+131259 & 0.952 & 0.0576 (0.0129) & 12.4 & K *  \\
150 & 174700.8+724026 & 0.986 & 0.3587 (0.0105) & 12.0 & G/K *+   \\
151 & 174704.1+332126 & 0.913 & 0.0541 (0.0097) & 12.6 & G/K *  \\
152 & 174947.6+335056 & 0.955 & 0.0645 (0.0105) & 12.3 & K/M *  \\
153 & 175140.9+730509 & 0.985 & 0.0508 (0.0044) & 12.0 & G *  \\
154 & 175318.5+213028 & 0.900 & 0.1757 (0.0175) & 0.0 & G/K *+   \\
155 & 175607.5+545536 & 0.932 & 0.2383 (0.0112) & 13.1 & Unclassifiable  \\
156 & 175633.1-014310 & 0.905 & 0.0502 (0.0136) & 11.9 & G *  \\
157 & 175711.2+224712 & 0.921 & 0.1116 (0.0154) & 13.3 & K/M *+   \\
158 & 175718.5+313314 & 0.976 & 0.1440 (0.0157) & 12.0 & G/K *  \\
159 & 175733.7+584414 & 0.968 & 0.0617 (0.0052) & 12.7 & K/M *+   \\
160 & 175758.9+550608 & 0.974 & 0.1184 (0.0085) & 12.4 & K/M *+   \\
161 & 175809.3+092241 & 0.968 & 0.1021 (0.0205) & 12.1 & G/K *  \\
162 & 175910.1+584300 & 0.992 & 0.0601 (0.0050) & 11.5 & G *+   \\
163 & 180147.5+273918 & 0.990 & 0.0721 (0.0115) & 10.8 & G *+   \\
164 & 180214.5+470112 & 0.968 & 0.0500 (0.0079) & 12.1 & G/K *   \\
165 & 180303.6+255932 & 0.939 & 0.0941 (0.0140) & 12.6 & K/M *+   \\
166 & 180305.8-033740 & 0.901 & 0.1102 (0.0201) & 13.1 & K/M *+   \\
167 & 180426.3+393044 & 0.969 & 0.0733 (0.0108) & 12.5 & G *  \\
168 & 180853.5+370702 & 0.907 & 0.0512 (0.0094) & 13.6 & G/K/M *  \\
169 & 181228.2+544701 & 0.930 & 0.0539 (0.0060) & 12.0 & F/G *  \\
170 & 181258.6+410604 & 0.908 & 0.0644 (0.0103) & 12.6 & K/M *  \\
171 & 181537.9+381927 & 0.905 & 0.0711 (0.0107) & 0.0 & ** (F/G * and K *)  \\
172 & 181610.9+585539 & 0.976 & 0.0542 (0.0052) & 12.0 & F/G *  \\
173 & 181616.5+541019 & 0.918 & 0.2920 (0.0129) & 13.8 & M *+  \\ 
174 & 181725.6+482202 & 0.970 & 0.1297 (0.0110) & 12.8 & M *+   \\
175 & 181937.8+364057 & 0.960 & 0.1496 (0.0148) & 13.0 & K/M *+   \\
176 & 182030.0+580437 & 0.929 & 0.8715 (0.0199) & 13.8 & Unclassifiable \\ 
\hline
\tablebreak
177 & 182247.1+443442 & 0.957 & 0.0853 (0.0099) & 13.1 & K/M *+    \\
178 & 182533.4+623416 & 0.982 & 0.1327 (0.0075) & 12.0 & K *+   \\
179 & 182539.1-021122 & 0.908 & 0.0505 (0.0139) & 13.1 & F/G *  \\
180 & 182903.6+435615 & 0.995 & 0.2351 (0.0173) & 8.3 & G *  \\
181 & 182931.3+223426 & 0.920 & 0.0613 (0.0110) & 12.4 & K * \\
182 & 183037.6+433555 & 0.956 & 0.0978 (0.0121) & 12.8 & K *+   \\
183 & 183159.2+054017 & 0.935 & 0.0870 (0.0169) & 11.9 & F/G *  \\
184 & 183219.2+021456 & 0.961 & 0.2539 (0.0273) & 12.3 & K/M *+   \\
185 & 183355.9+514313 & 0.962 & 2.4140 (0.0479) & 9.6 & K * + \\ 
186 & 183544.4+300808 & 0.913 & 0.0589 (0.0107) & 12.8 & K *+   \\
187 & 183824.9+340642 & 0.969 & 0.0945 (0.0144) & 12.2 & G *  \\
188 & 184632.2+485443 & 0.921 & 0.0680 (0.0101) & 13.2 & G/K *+   \\
189 & 184640.1-091622 & 0.958 & 0.1494 (0.0251) & 12.7 & M *+   \\
190 & 184653.6+321652 & 0.937 & 0.0550 (0.0096) & 12.2 & G *  \\
191 & 184725.7+084106 & 0.937 & 0.0664 (0.0166) & 12.9 & K/M *+   \\
192 & 184739.3+015732 & 0.942 & 0.7512 (0.0477) & 12.8 & Unclassifiable \\
193 & 184752.3+275703 & 0.932 & 0.0818 (0.0126) & 13.0 & K/M *+   \\
194 & 184830.1+642330 & 0.966 & 0.0985 (0.0058) & 12.6 & G/K *  \\
195 & 184847.1+804022 & 0.995 & 0.1766 (0.0136) & 11.1 & G *  \\
\enddata

\tablecomments{Notation: *=star; **=double; +=coronally active;}

\end{deluxetable}

\clearpage

\begin{deluxetable}{cccc}
\tabletypesize{\scriptsize}
\tablecaption{\label {tab:prev} Table of Previous Classifications}
\tablewidth{13cm}
\tablehead{
\colhead{\#} &
\colhead{Source Name} &
\colhead{Source Class} &
\colhead{Previous Classification} \\
\colhead{} &
\colhead{(1 RXS J)} &
\colhead{} &
\colhead{}
}

\startdata
14 & 111232.8+354845 & G * & HiPM* (G0V type)$^1$ \\
28 & 115107.6+351617 & M *  & HiPM* (M1 type)$^2$\\
33 & 121506.7+331129 & Unclassifiable & AGN$^1$\\
37 & 122212.6+731453 & G *+  & Variable RSCVn * (G7V type)$^1$\\
44 & 125534.6+270355 & ** (K/M *+ and *+)  & M0e *$^3$\\
48 & 130346.2+283729 & G *+  & Variable RSCVn * (G8V type)$^4$\\
50 & 131152.9+275242 & Saturated & HiPM* (G0V type)$^5$\\
56 & 133443.7-082036 & K *+   & Flare * (K5Ve type)$^6$\\
70 & 143440.3+294441 & Saturated  & V * (F2V type)$^7$\\
83 & 151750.8+050615 & AGN/Galaxy  & AGN$^8$\\
85 & 152244.2+161550 & G *  & Eclipsing ** of W UMa type (G5 type)$^1$\\
117 & 165601.2+650807 & F *  & Spectroscopic ** (F6Vvar type)$^1$\\
128 & 171632.8+430221 & K/M * & Galaxy$^7$\\
173 & 181616.5+541019 & M *+  & M dwarf * (KV:e type)$^9$\\
176 & 182030.0+580437 & Unclassifiable & White dwarf (DA type)$^{10} $\\
181 & 182931.3+223426 & K * & Eclipsing ** of Algol type$^{11}$\\
185 & 183355.9+514313 & K *+ & V* of BY DRA type (K6Ve type)$^{12}$\\
192 & 184739.3+015732 & Unclassifiable & DA (white dwarf)$^{10}$\\
\enddata

\tablerefs{$^1$\citet{RosatCat};
$^2$\citet{Gliese};
$^3$\citet{RosatCat2};
$^4$\citet{Vilnius}; 
$^5$\citet{NEXXUS}; 
$^6$\citet{Systematic03}; 
$^7$\citet{Hamburg03};
$^8$\citet{QSO}; 
$^9$\citet{MDwarf};
$^{10}$\citet{Hardxray04}; 
$^{11}$\citet{VarList}; 
$^{12}$\citet{Study};
}
\tablecomments{Notation: *=star; **=double; +=coronally active;
 V*=Variable Star; HiPM*=High proper motion star;}

\end{deluxetable}

\begin{deluxetable}{ccc}
\tabletypesize{\scriptsize}
\tablecaption{\label {tab:comments} Comments on Individual Sources}
\tablewidth{15cm}
\tablehead{
\colhead{\#} &
\colhead{Source Name} &
\colhead{Comments} \\
\colhead{} &
\colhead{(1 RXS J)} &
\colhead{} 
}

\startdata
3 & 103945.8+654529 & Balmer emission \\
8 & 105541.8+424603 & several lines are double \\
9 & 105634.4+583535 & some lines are double \\
10 & 105706.3-010109 & {\boldit on spectrum of G/K star:}  {\bf no} CaH\&K \\
16 & 111612.6+494208 & Balmer emission \\
22 & 113316.5+134427 & Balmer emission (strong H${\alpha}$) \\
23 & 113336.8+075131 & Balmer emission \\
24 & 113352.8+361331 & Balmer emission \\
25 & 113533.7+825920 & Balmer emission \\
29 & 115651.8+082726 & Balmer emission \\
30 & 115938.3+560301 & Balmer emission, inconspicuous CaH\&K \\
31 & 120732.6+471447 & inconspicuous CaH\&K \\
32 & 121215.7+685306 & Balmer emission (strong H${\alpha}$) \\
34 & 121754.3+105647 & inconspicuous CaH\&K \\
37 & 122212.6+731453 & CaH\&K with double core emission; most lines are double \\
38 & 122308.2+452801 & weak Balmer emission \\
44 & 125534.6+270355 & {\boldit on spectrum of K/M star:} most lines are double \\
 & & {\boldit on spectrum of unclassified star:} Balmer emission \\
48 & 130346.2+283729 & Balmer emission \\
51 & 132117.1+210116 & H Balmer lines difficult to find \\
52 & 132641.4+022230 & inconspicuous CaH\&K \\
53 & 132713.0+455826 & strong Balmer emission, He I D3 line in emission; broad lines \\
55 & 132931.8+514214 & Balmer emission, inconspicuous CaH\&K \\
56 & 133443.7-082036 & Balmer emission, CaH\&K with strong core emission \\
57 & 133740.0+320930 & Balmer emission \\
59 & 134838.5+712552 & weak Balmer emission, Ca I line difficult to find, inconspicuous CaH\&K \\
61 & 135510.3+160837 & Balmer emission \\
62 & 135728.9+591107 & inconspicuous CaH\&K \\
63 & 135902.2+591754 & Balmer emission \\
65 & 141508.0+124249 & H Balmer lines difficult to find, inconspicuous CaH\&K \\
66 & 141630.7+265523 & H${\alpha}$ line with P Cygni profile; most other lines are double with core emission for CaK \\
78 & 144952.3+420615 & H Balmer lines difficult to find \\
79 & 145502.5+144210 & Balmer emission \\
81 & 150507.5+573036 & inconspicuous CaH\&K \\
82 & 151145.7+101421 & Balmer emission \\
84 & 151948.2+070754 & most lines are double \\
87 & 152757.4+651533 & inconspicuous CaH\&K \\
88 & 153726.3+362343 & H Balmer lines difficult to find, inconspicuous CaH\&K \\
89 & 154058.3+402657 & inconspicuous CaH\&K \\
91 & 154355.8+132555 & inconspicuous CaH\&K \\
93 & 154623.6+440318 & Balmer emission, inconspicuous CaH\&K \\
94 & 154718.8+631101 & Balmer emission, inconspicuous CaH\&K \\
95 & 154809.0+161207 & Balmer emission; some lines are double \\
\hline
\tablebreak
96 & 154925.0+460822 & {\boldit on spectrum of G/K star:}  weak Balmer emission, inconspicuous CaH\&K \\
 & & {\boldit on spectrum of K/M star:}  inconspicuous CaH\&K \\
97 & 155752.5+774738 & inconspicuous CaH\&K \\
98 & 155842.6+323047 & Balmer emission, inconspicuous CaH \\
99 & 155947.5+440358 & Balmer emission, inconspicuous CaH \\
100 & 160157.0+182514 & weak Balmer emission \\
101 & 160248.3+252031 & CaH\&K with strong core emission \\
102 & 160438.3+702212 & Balmer emission \\
103 & 160524.8-085549 & Balmer emission \\
104 & 160704.4+023823 & CaK with double core emission \\
106 & 161712.7+773345 & Balmer emission, inconspicuous CaH \\
107 & 162506.5+300218 & Balmer emission, CaH\&K with strong core emission \\
109 & 162708.9+662620 & Balmer emission, inconspicuous CaH \\
111 & 163259.1+504433 & Balmer emission, inconspicuous CaH \\
113 & 163741.2+291946 & Balmer emission; broad lines \\
114 & 164212.9+655300 & Balmer emission, CaK difficult to find \\
115 & 165315.4+701554 & inconspicuous CaH\&K; H ${\alpha}$ blueshifted by approx. 2 ${\AA}$ \\
118 & 165733.5+593156 & H ${\alpha}$ blueshifted by approx. 2 ${\AA}$ \\
119 & 165909.5+205807 & Balmer emission \\
120 & 170157.5+073329 & inconspicuous CaH; H ${\alpha}$ blueshifted by approx. 2 ${\AA}$ \\
121 & 170248.8+471258 & Balmer emission \\
122 & 170420.7+392909 & Balmer emission, inconspicuous CaH \\
123 & 170757.4+291922 & strong Balmer emission \\
124 & 171017.5+632135 & inconspicuous CaH\&K \\
125 & 171206.0+454025 & inconspicuous CaH \\
127 & 171355.0+455955 & inconspicuous CaH\&K \\
128 & 171632.8+430221 & inconspicuous CaH\&K \\
129 & 171800.1+212816 & some lines with core emission \\
130 & 172103.4+114237 & {\boldit on spectrum of M star:}  weak Balmer emission \\
131 & 172128.0+084446 & weak Balmer emission \\
132 & 172157.8+574913 & Balmer emission \\
134 & 172339.8+352757 & inconspicuous CaH\&K \\
135 & 172413.5+402616 & inconspicuous CaH\&K \\
136 & 172534.0+160923 & strong Balmer emission \\
137 & 172812.2+723923 & Balmer emission \\
138 & 172927.2+352402 & Balmer emission, inconspicuous CaH; H ${\alpha}$ blueshifted by approx. 2 ${\AA}$ \\
139 & 173039.4+785554 & Balmer emission, inconspicuous CaH \\
143 & 173734.1+414618 & Balmer emission, inconspicuous CaH\&K \\
144 & 173748.1+215925 & Balmer emission, CaK difficult to find \\
145 & 173855.7+263400 & Balmer emission \\
146 & 174120.9+084312 & strong Balmer emission \\
147 & 174311.5+334950 & Balmer emission \\
148 & 174411.0+720304 & CaH\&K with strong core emission \\
150 & 174700.8+724026 & strong Balmer emission, CaH\&K strong core emission \\
\hline
\tablebreak
151 & 174704.1+332126 & inconspicuous CaH \\
152 & 174947.6+335056 & H Balmer lines difficult to find, inconspicuous CaH\&K \\
154 & 175318.5+213028 & Balmer emission; H ${\alpha}$ blueshifted by approx. 2 ${\AA}$ \\
156 & 175633.1-014310 & inconspicuous CaH\&K \\
157 & 175711.2+224712 & Balmer emission \\
158 & 175718.5+313314 & inconspicuous CaH\&K \\
159 & 175733.7+584414 & Balmer emission \\
160 & 175758.9+550608 & Balmer emission, CaH\&K with strong core emission \\
161 & 175809.3+092241 & CaK difficult to find \\
162 & 175910.1+584300 & CaH\&K with double core emission; most lines are double \\
163 & 180147.5+273918 & CaH\&K with core emission; most lines are double \\
164 & 180214.5+470112 & most lines are double or broad; H ${\alpha}$ blueshifted by approx. 3 ${\AA}$ \\
165 & 180303.6+255932 & H Balmer lines difficult to find; several lines with core emission \\
166 & 180305.8-033740 & Balmer emission \\
167 & 180426.3+393044 & H Balmer lines difficult to find, CaK difficult to find \\
168 & 180853.5+370702 & inconspicuous CaH\&K \\
170 & 181258.6+410604 & inconspicuous CaH\&K \\
173 & 181616.5+541019 & strong Balmer emission, CaH\&K with strong core emission \\
174 & 181725.6+482202 & Balmer emission, CaH\&K with strong core emission \\
175 & 181937.8+364057 & Balmer emission, CaH\&K with strong core emission \\
177 & 182247.1+443442 & Balmer emission, CaH\&K with strong core emission \\
178 & 182533.4+623416 & CaH\&K with strong core emission \\
179 & 182539.1-021122 & H Balmer lines difficult to find, inconspicuous CaH\&K \\
182 & 183037.6+433555 & CaH\&K with core emission \\
183 & 183159.2+054017 & broad lines \\
184 & 183219.2+021456 & strong Balmer emission, CaH\&K with strong core emission \\
185 & 183355.9+514313 & Balmer emission, CaH\&K with strong core emission \\
186 & 183544.4+300808 & Balmer emission \\
188 & 184632.2+485443 & Balmer emission \\
189 & 184640.1-091622 & Balmer lines double and in emission, inconspicuous CaH\&K \\
191 & 184725.7+084106 & Balmer emission \\
193 & 184752.3+275703 & Balmer emission \\
\enddata

\tablecomments{Notation: Balmer emission=H Balmer lines in emission; CaH/CaK/CaH\&K= Ca II H line or Ca II K line or both Ca II H and K lines; inconspicuous CaH/CaK/CaH\&K=noisy spectrum with Ca II H line or Ca II K line or both Ca II H and K lines difficult to find;}

\tablecomments{For object 171632.8+430221 (\# 128) our classification differs from the previous classification as a galaxy  \citep{Hamburg03}.
 Our conclusion is based on
 spectral features which indicate that the optical counterpart is a K
 or M type star, and on a cross-check with the 2MASS image database
 which showed that the counterpart has a narrow point spread
 function, indicating a point-source. This discrepancy may also be due to the fact that we have not observed
the same optical counterpart as previously \citep{Hamburg03}.}

\end{deluxetable}

\clearpage

\begin{figure} 
\caption{ \label{fig:1} H${\alpha}$ line at 6566 {\AA} for 1RXS J154150.8+310031 (object \# 90) }
\end{figure}

\begin{figure} 
\caption{ \label{fig:2} H${\alpha}$ line at 6562 {\AA} for 1RXS J165601.2+650807 (object \# 117) }
\end{figure}

\begin{figure} 
\caption{ \label{fig:3} TiO lines at 7088.2 {\AA} and 7123.9 {\AA} for 1RXS J181725.6+482202 (object \# 174) }
\end{figure}

\begin{figure} 
\caption{ \label{fig:4} H${\alpha}$ line in emission at 6561.7 {\AA} for 1RXS J181725.6+482202 (object \# 174) }
\end{figure}

\begin{figure} 
\caption{ \label{fig:5} H${\alpha}$ line with P Cygni profile at 6561.9 {\AA} for 1RXS J141630.7+265523 (object \# 66) }
\end{figure}

\begin{figure} 
\caption{ \label{fig:6} Na I D with double lines at 5889.8 {\AA} and 5896.3 {\AA} for 1RXS J141630.7+265523 (object \# 66) }
\end{figure}

\begin{figure} 
\caption{ \label{fig:7} Ca II H line with core emission at 3967.8 {\AA} for 1RXS J113858.9+421957 (object \# 26) }
\end{figure}

\pagestyle{empty}
\begin{figure}[htb]
\PSbox{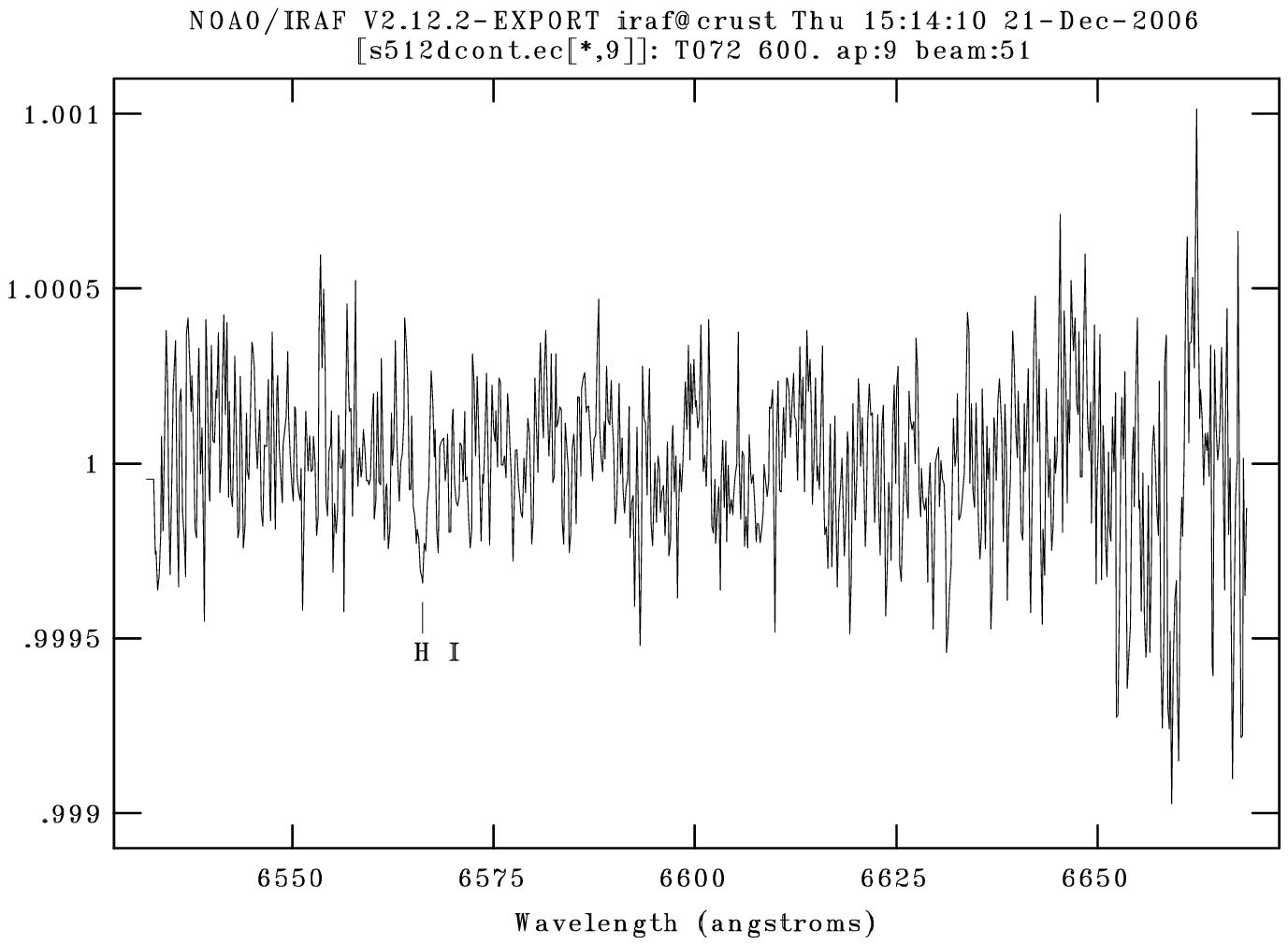 hoffset=-80 voffset=-240}{14.7cm}{21.5cm}
\FigNum{\ref{fig:1}}
\end{figure}
\clearpage

\pagestyle{empty}
\begin{figure}[htb]
\PSbox{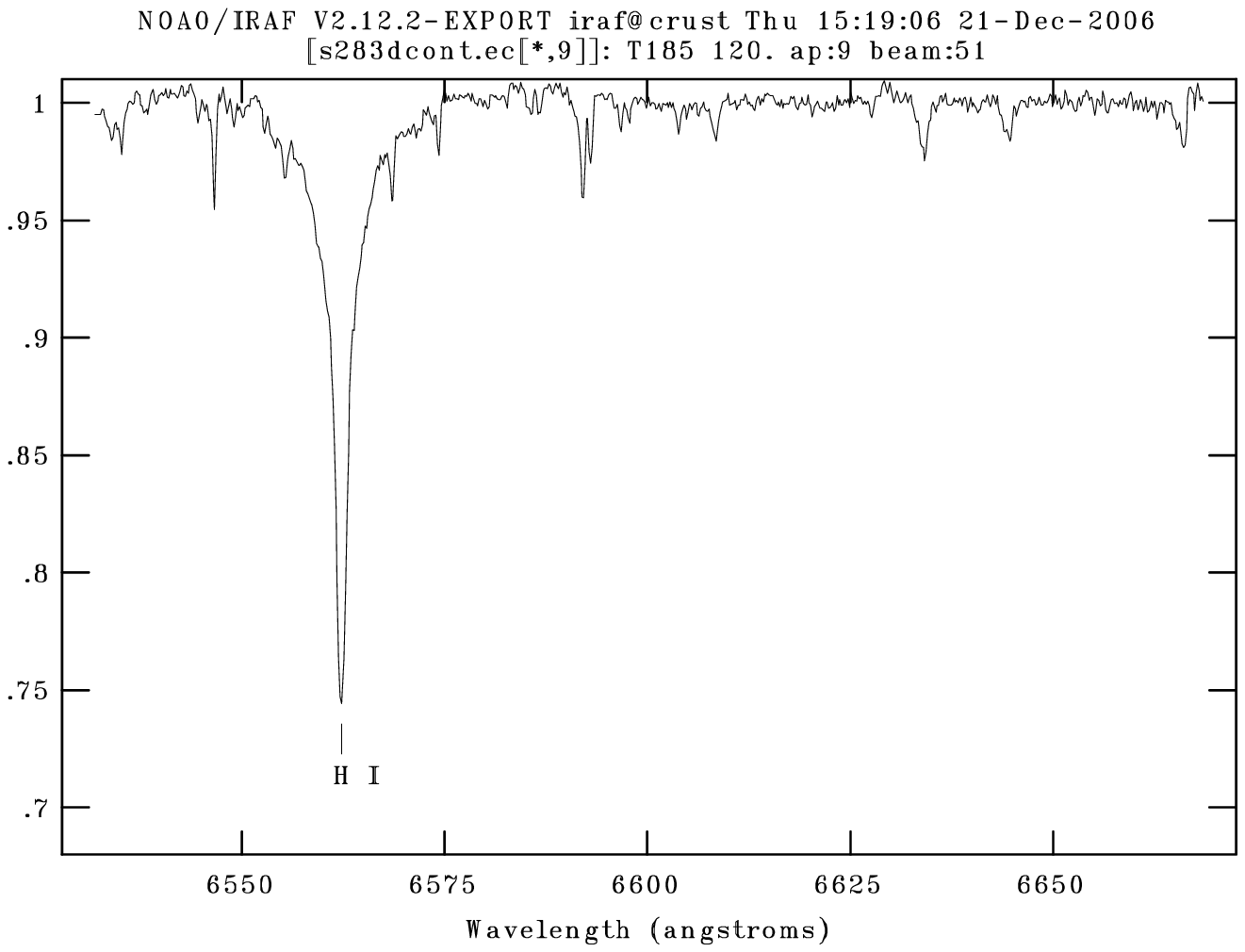 hoffset=-80 voffset=-240}{14.7cm}{21.5cm}
\FigNum{\ref{fig:2}}
\end{figure}
\clearpage

\pagestyle{empty}
\begin{figure}[htb]
\PSbox{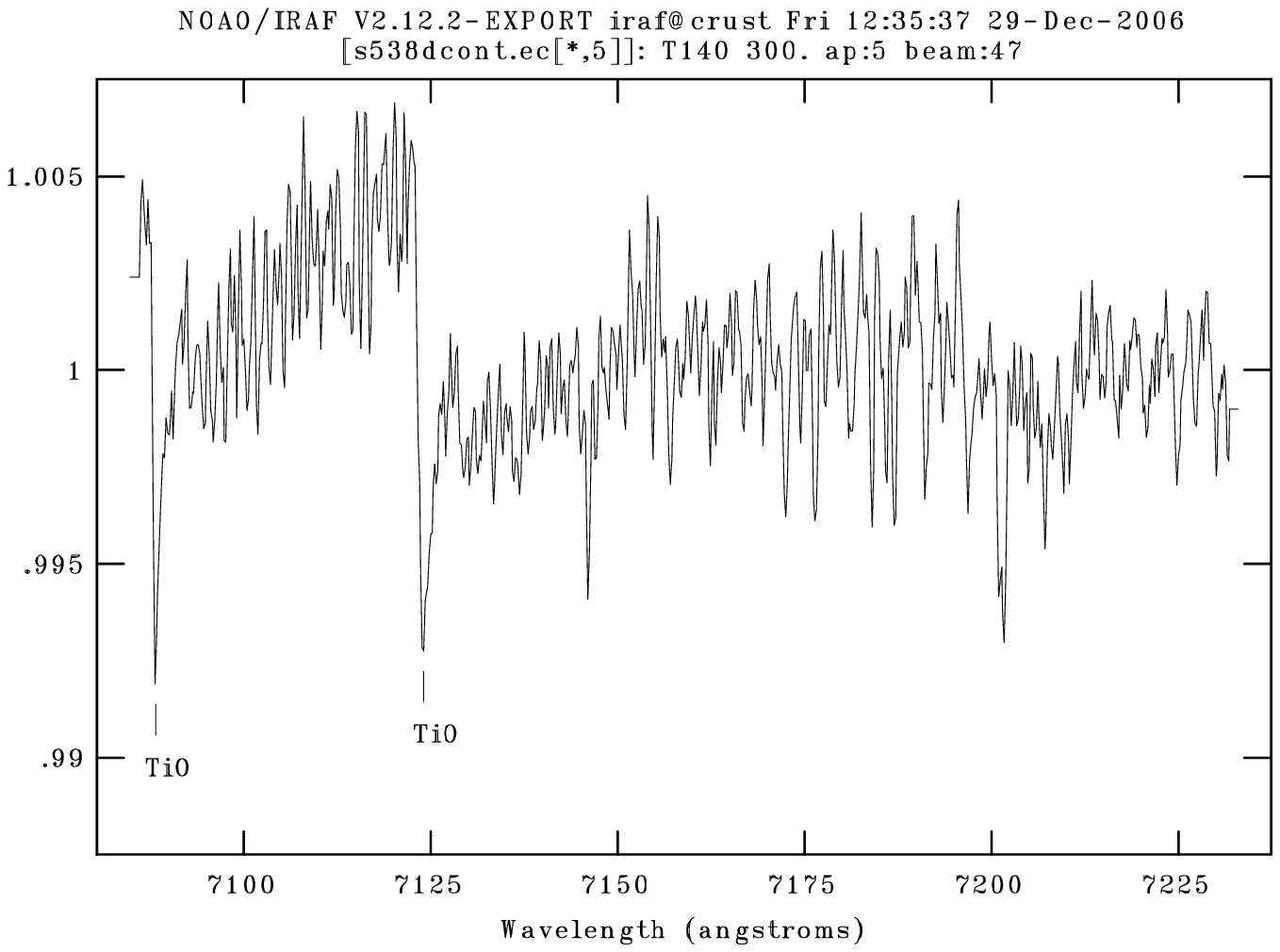 hoffset=-80 voffset=-240}{14.7cm}{21.5cm}
\FigNum{\ref{fig:3}}
\end{figure}
\clearpage

\pagestyle{empty}
\begin{figure}[htb]
\PSbox{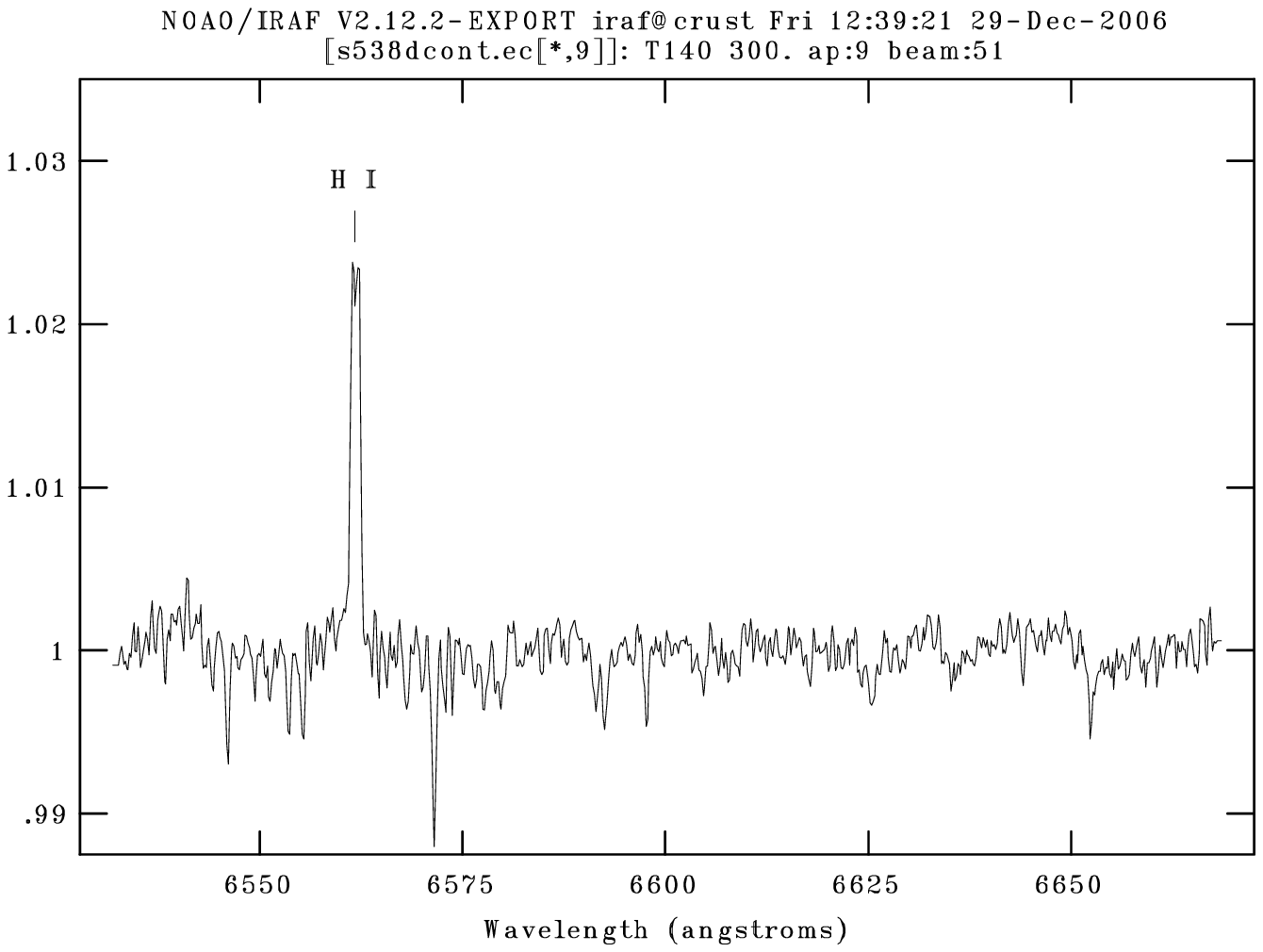 hoffset=-80 voffset=-240}{14.7cm}{21.5cm}
\FigNum{\ref{fig:4}}
\end{figure}
\clearpage

\pagestyle{empty}
\begin{figure}[htb]
\PSbox{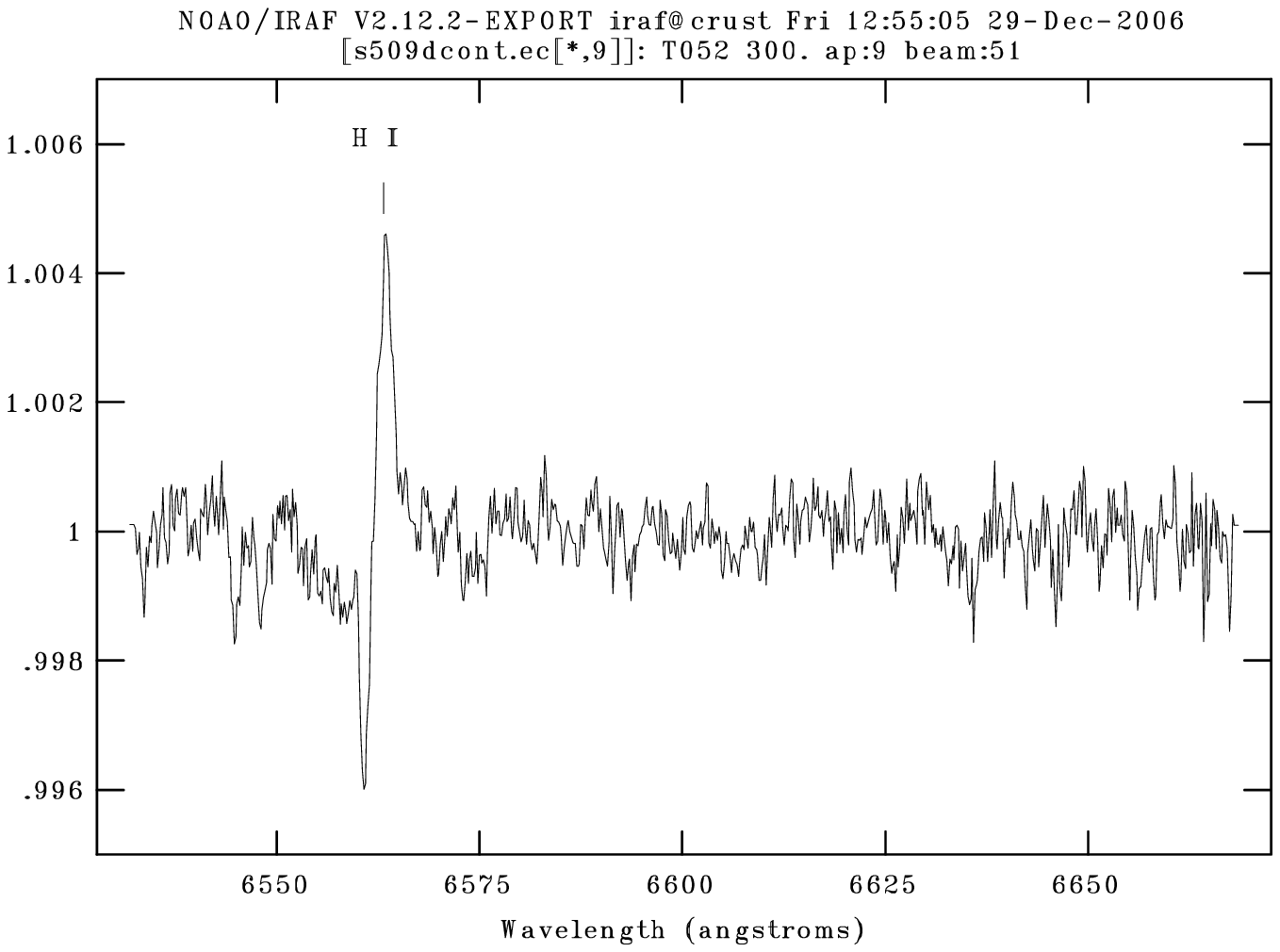 hoffset=-80 voffset=-240}{14.7cm}{21.5cm}
\FigNum{\ref{fig:5}}
\end{figure}
\clearpage

\pagestyle{empty}
\begin{figure}[htb]
\PSbox{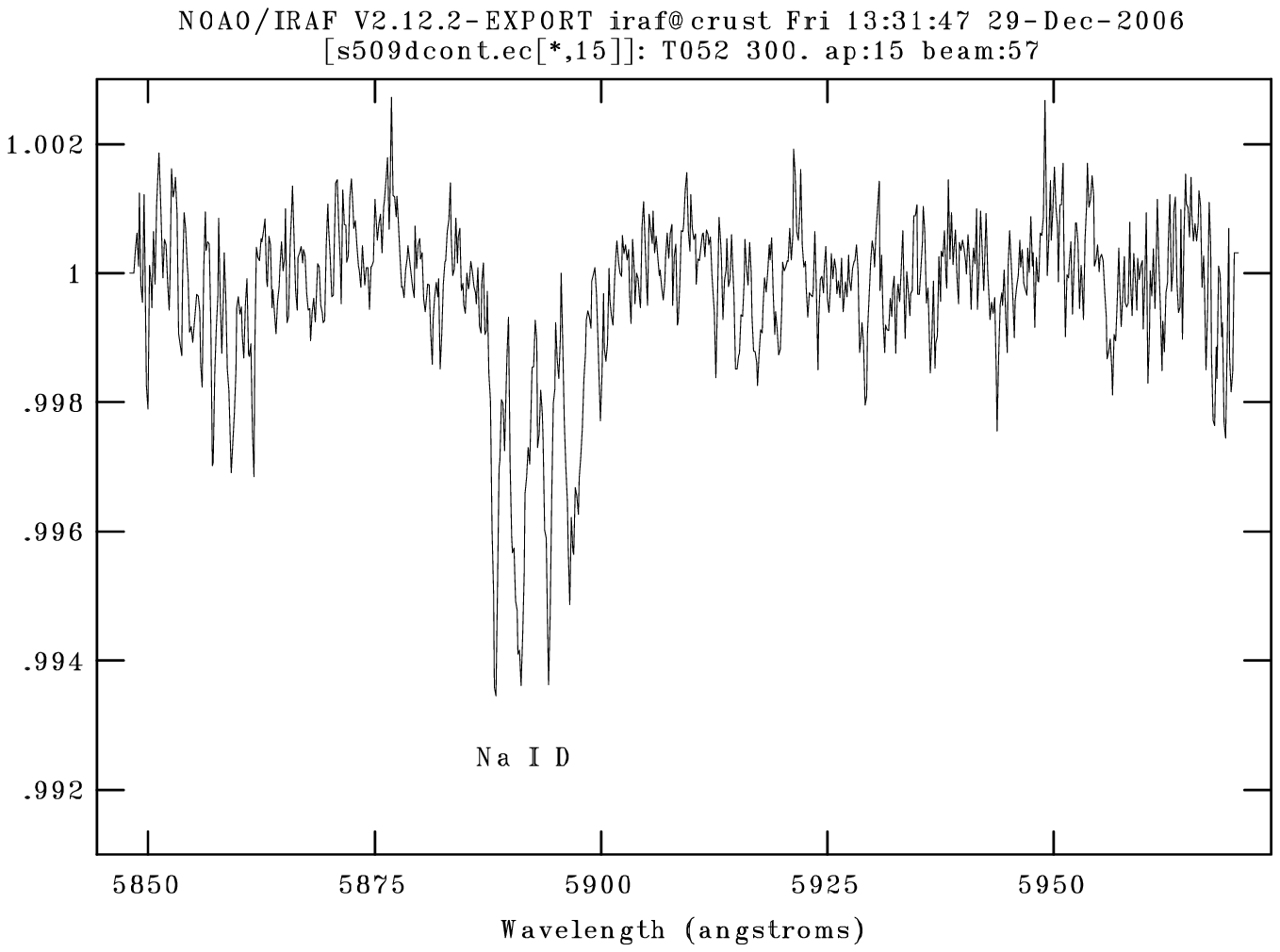 hoffset=-80 voffset=-240}{14.7cm}{21.5cm}
\FigNum{\ref{fig:6}}
\end{figure}
\clearpage

\pagestyle{empty}
\begin{figure}[htb]
\PSbox{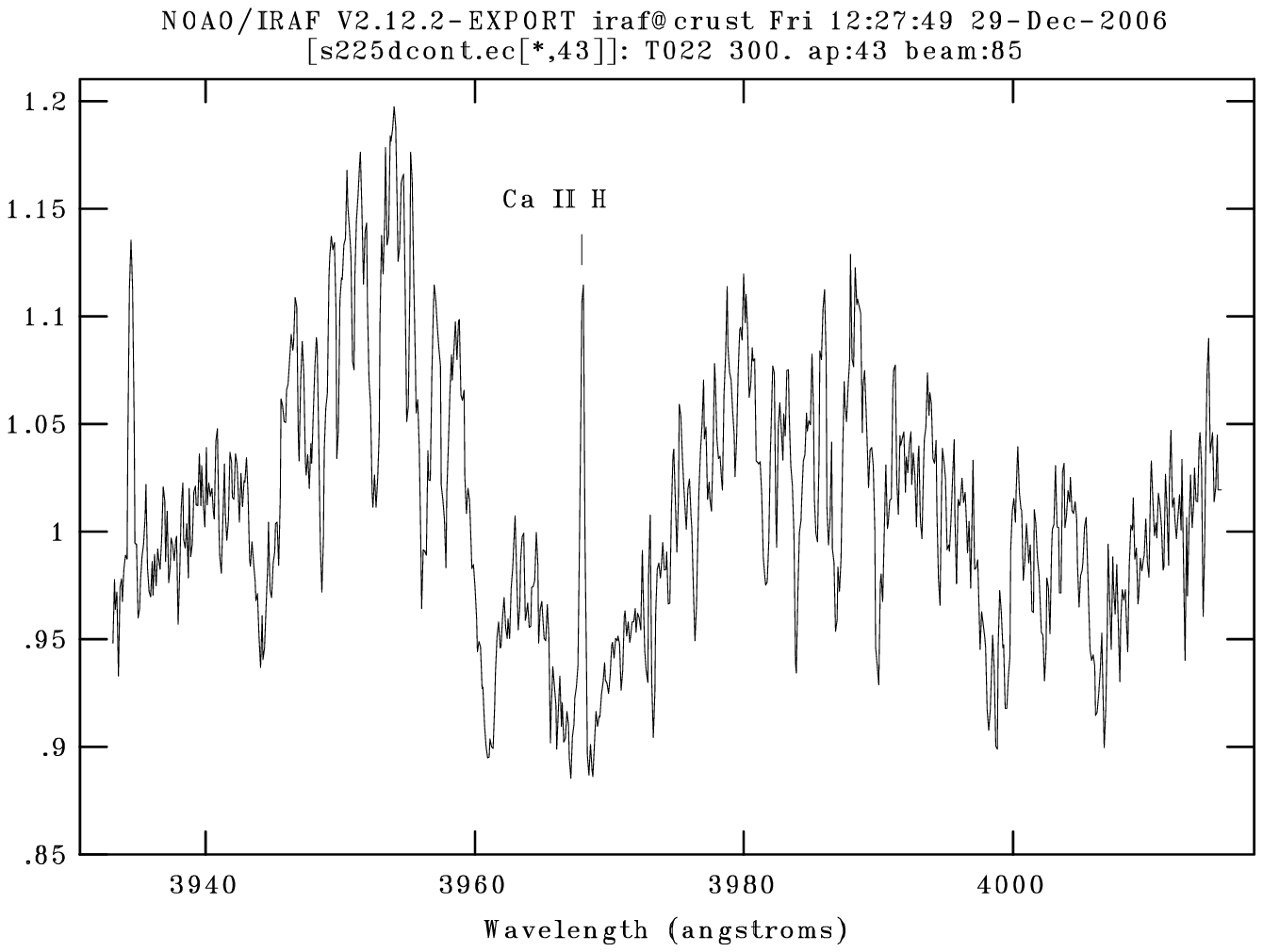 hoffset=-80 voffset=-240}{14.7cm}{21.5cm}
\FigNum{\ref{fig:7}}
\end{figure}
\clearpage

\end{document}